# Tumor can originate from not only rare cancer stem cells

Min Hu[1], Yu-Fei He[2*]

[1]Tongji Hospital, Tongji Medical College, Huazhong University of Science and Technology, Wuhan, P.R. China.

[2] Liver Cancer Institute and Institute of Biomedical Sciences, Fudan University, Shanghai, P.R. China.

* To whom correspondence should be addressed. Present address: Institute of Biochemistry and Cell Biology, Chinese Academy of Sciences, 320 Yueyang Road, Shanghai 200031, P.R. China

Tel.: 86-21-54921334; Fax: +86-21-54921226. E-mail address: heyufei@fudan.edu.cn.

**Abstract**

Tumors are believed to consist of a heterogeneous population of tumor cells originating from rare cancer stem cells (CSCs). However, emerging evidences show that tumor may also originate from non-CSCs. Here, we give evidences supporting that the number of tumorigenic tumor cells is higher than the number of CSCs and tumor can also derive from non-CSCs. First, we applied an idealized mathematical model and theoretically calculated that non-CSCs could initiate tumor if their proliferation potential was adequate. Next, we demonstrated by experimental studies that 17.7%, 38.6% and 5.2% of tumor cells in murine B16 solid melanoma, H22 hepatoma and Lewis lung carcinoma, respectively, were potentially tumorigenic. We propose that the rare CSCs, if exist, are not the only origination of a tumor.

## 1. Introduction

Tumors are comprised of phenotypically diverse populations of tumor cells, and some of them are highly tumorigenic than others (1). It is now widely believed that many of cancers are derived from a rare subpopulation of cells, designated cancer stem cells (CSCs), or cancer-initiating cells (2). By definition, cancer stem cells are self-renewing cells that are responsible for sustaining a cancer and for producing differentiated progeny that form the bulk of the cancer, while cancer-initiating cell is a general term that encompasses both cancer cell of origin (precancerous cell that gives rise to a cancer stem cell) and cancer stem cell (3). However, CSCs always have the same meaning as cancer-initiating cells in many literatures (2, 4). The concept of CSC has been demonstrated in many human cancers including melanoma (5), lung (6), liver (7), and colon caners (4), and so on. Like stem cells from their normal counterpart, CSCs (or cancer-initiating cells) are a rare population, and they are self-renewing and capable of unlimited proliferation. Tumors are believed to be generated from such rare CSCs (1, 2). If the CSC hypothesis is correct, it will usher in an era of hope for curing once-incurable cancers by targeting CSC. However, not everyone accepts the hypothesis of CSC.

Recently a paper challenged the concept of CSC because the authors concluded that tumor growth needed not be driven by rare CSC (8). Using nonirradiated histocompatible recipient mice (not sublethally irradiated nonobese diabetic severe combined immunodeficient mice that usually were used in the studies of CSC identification), the authors observed that a large proportion of primary tumor cells can sustain the growth of murine lymphoid and myeloid malignancies. More recently, a study showed that in single-cell transplants, an average of 27% of unselected melanoma cells from four different patients could form tumors in highly immunocompromised mice, and thus the frequency of tumorigenic cells in human melanoma was much higher than reported for any cancer previously suggested to follow a cancer stem-cell model (9).

Thus, the concept of CSC seems somewhat controversial. Is the concept of CSC correct that the tumor growth need just be driven by rare CSC, or the tumor growth need not be driven just by CSC? In the present study, we provide evidences supporting that the number of tumor cells capable of generating a new tumor is higher than the number of CSCs. We give the definition of tumorigenic cells as the tumor cells capable of generating a new tumor. Tumorigenic cells encompass both CSCs and others tumor cells that can generate a new tumor *in vivo*. Although the CSCs might exist, they might not be the only ones to generate tumors. Therefore, during tumor therapy, it seems necessary to target all tumor cells that have a high proliferation potential and not only CSCs.

## 2. Material and methods
### *2.1. Mathematical model and calculation*

The theoretical probability of CSC in a population of tumor cells was calculated by using traditional mathematic tools. To easily understand and calculate, we also used an idealize CSC differentiation model (Fig. 1).

### *2.2. Mice and Tumor Models*

Animal experiments involving mice were approved by the institute's Animal Care and Use

Committee. Mice were maintained under standard conditions. To generate solid melanoma, C57BL/6 mice were inoculated with B16 melanoma cells by the subcutaneous injection of $2\times10^6$ cells into the dorsal skin. To generate solid hepatoma, BALB/c mice were inoculated with H22 hepatoma cells by the injection of $5\times10^5$ cells into the hind thigh muscle. To generate solid lung cancer, C57BL/6 mice were inoculated with Lewis lung carcinoma (LLC) cells by the axillary injection of $2\times10^6$ cells.

### *2.3. Single tumor cell preparation*

Tumors (1 cm in diameter) were minced into 2–3-mm pieces and digested for 2 h at 37°C in DMEM containing type IV collagenase (0.1%; Sigma), type I DNase (0.002%; Sigma), and type V hyaluronidase (0.01%; Sigma) and single-cell suspension was prepared. The cells were then washed twice with PBS for 5 min at 120 g.

Further removing of blood cells and dead cells by magnetic activated cell-sorting was performed as described previously (10). Briefly, to remove leukocytes and erythrocytes, biotinylated anti-CD45.2 and anti-TER119 monoclonal antibodies (eBioscience) were used. To remove dead cells, we used Dead Cell Removal MicroBeads (Milteny Biotec) following the manufacture's protocol. After the antibody labeling, cells were resuspended in 80 μl of buffer (PBS containing 0.5% BSA and 2 mM EDTA) per $10^7$ total cells and incubated with anti-biotin microbeads for 15 min at 6-12 °C. Cells were washed twice and finally resuspended in 500 μl of buffer per $10^7$ total cells. A pre-moistened MS column (Milteny Biotec) was placed in the magnetic field of a suitable MACS separator. The cell suspension was applied onto the column and effluents were collected as the purified tumor cells. Viability of sorted tumor cells exceeded 98% as assessed by trypan blue exclusion.

### *2.4. In vitro colony assay*

Purified tumor cells were plated on 96-well plates at a density of 1 cell per well and maintained at 37°C, 5% $CO_2$. The medium was DMEM supplemented with 15% FCS and 100 U/ml penicillin-streptomycin. Two weeks later, the large colonies were digested (if B16 and LLC cells) and transferred to new culture dishes.

### *2.5. Tumor cell inoculation and tumor surveillance*

Mice were inoculated with different doses of tumor cells in 100 μl of PBS. Tumor surveillance was performed as described previously (11). Tumor size was measured using calipers fitted with a Vernier scale when tumor could be palpated. The tumor diameter was calculated using the formula ($a + b$)/2, with $a$ as the larger diameter and $b$ as the smaller diameter. Mice were sacrificed 100 days after inoculation or for ethical reasons when tumors exceeded 1.2 cm in diameter. All tumor tissues were examined and verified by hematoxylin-eosin staining.

## 3. Results
### *3.1. Theoretical calculation of CSC probability*

CSCs are a very rare population in the tumor (1, 2). For example, there is only one cancer-initiating cell in $5.7\times10^4$ unfractionated human colon tumor cells (4). According to the CSC theory, the tumor formation depends on the existence of CSC, or cancer is maintained by CSC. In other words, if a small tumor cell population taken from a tumor contains at least one CSC, then, theoretically, a new tumor could be generated from such population of tumor cells. If there are $10^n$ CSCs in a 1-$cm^3$ tumor mass which contains approximately $10^9$ tumor cells, then the probability that there is at least one CSC in any $10^m$ tumor cells taken from such $10^9$ tumor cells can be calculated using binomial distribution (12) as follows.

The probability that there is no CSC in any $10^m$ tumor cells taken from such $10^9$ tumor cells is:

$$p' = \frac{\dfrac{(10^9 - 10^n)!}{(10^9 - 10^n - 10^m)!(10^m)!}}{\dfrac{(10^9)!}{(10^9 - 10^m)!(10^m)!}} \times 100\%$$

Then, the probability that there is at least 1 CSC in any $10^m$ tumor cells taken from the $10^9$ tumor cells is:

$$p = (1 - p') \times 100\%$$

That is,

$$P = \left[1 - \frac{(10^9 - 10^n)!(10^9 - 10^m)!}{(10^9 - 10^n - 10^m)!(10^9)!}\right] \times 100\%$$

Based on the past experience of tumor inoculation, nearly 100% of mice inoculated with any $10^5$ (or more) unsorted tumor cells would develop a tumor. By applying the above formula, we calculate that there are about $10^4$ CSCs in a 1-cm$^3$ tumor mass, i.e., the proportion of CSC in the tumor is about $10^{-5}$, which is a value close to that obtained from the experimental study of human colon cancer (1 CSC in about $5.7 \times 10^4$ colon cancer cells (4)). Theoretically, if the proportion of CSC is 1%, then the probability that there is at least one CSC in any 100 tumor cells taken from a 1-cm$^3$ tumor mass is 0.63 (63%). The observations of experiments involving inoculation with tumor cells reveal that any 100 tumor cells from a 1-cm3 tumor mass cannot usually generate a new tumor; thus, the actual proportion of CSCs in a tumor is below 1%.

Theoretically, a 1-cm$^3$ tumor mass could be generated from one CSC after 30 cell doublings ($2^{30} \approx 10^9$, Fig. 1). If the proportion of CSC is $10^{-5}$, the theoretical division times to generate a 1-cm$^3$ tumor mass should be about 17 ($10^9 \approx 10^4 \times 2^{17}$).

Although a CSC (the zero differentiation level, Fig. 1) is self-renewing and capable of unlimited proliferation, the proliferation potential of CSCs after 1 passage (the 1st differentiation level, Fig. 1) should be determined. Theoretically, if cells at the 1st differentiation level can divide 30 times, then 1 such cell can generate a 1-cm$^3$ new tumor mass after *in vivo* inoculation ($2^{30} \approx 10^9$). Further, if the cells at the 1st differentiation level had a higher proliferation potential (for example, 32 divisions), then the cells at the 2nd and 3rd differentiation levels would have the potential to generate a 1-cm$^3$ tumor mass after *in vivo* inoculation. If the cells at the 1st differentiation level in the model in Fig. 1 were capable of dividing 30 + y times, then the proportion of the cells in a 1-cm$^3$ tumor mass that could generate a 1-cm$^3$ new tumor is $2^{-30} + 2^{-30} + 2^{-29} + \ldots + 2^{y-30}$, i.e., $2^{y-29}$ (y ≤ 29). Taken together, the higher the proliferation potential (>30 divisions) of the cells at the 1st differentiation level, the higher the proportion of cells with the potential to generate a 1-cm$^3$ new tumor in a cell population. If y = 23, then theoretically, according to the model in Fig. 1, there will be more than 1% (i.e., $2^{y-29} = 2^{-6} = 0.016$) of tumor cells with the potential to generate a 1-cm$^3$ new tumor. Meanwhile, the proportion of CSCs was merely $2^{-30}$. In other words, the number of tumor cells capable of generating a new tumor is higher than the number of CSCs. Interestingly, such theoretical speculation conforms to the results of two recent studies (8, 9).

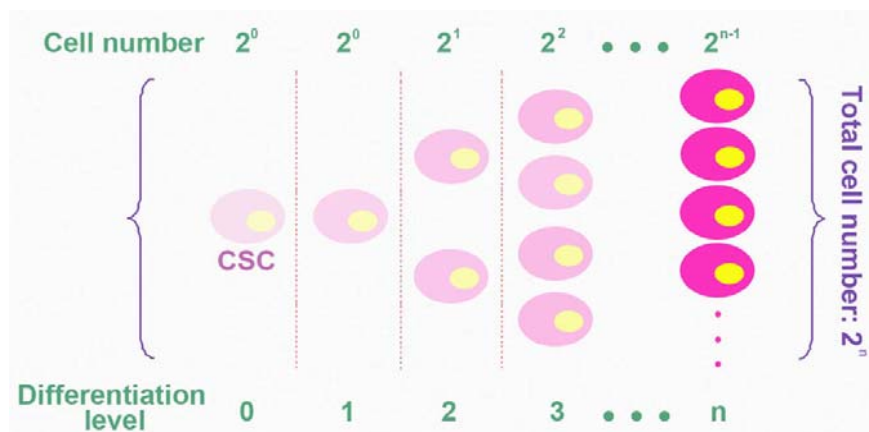

**Figure 1. An idealized model of CSC differentiation.** In fact, this model is not new and is well accepted by many people. In this model, CSCs (differentiation level 0) divide asymmetrically to self-renew and gives rise to a more differentiated progeny (differentiation level 1) that divide symmetrically and give rise to 2 more differentiated tumor cells. After n divisions (differentiation level n is the terminal differentiation level), the total number of cells derived from a single CSC would be $2^n$.

Of course, we should note that biologically a growing cell population may yield intra-tumoral quiescence (13, 14) and hence sub-exponential growth and the final tumor size should be smaller than the theoretical size. Thus, to generate a 1-cm$^3$ new tumor *in vivo*, actually one stem cell should divide much more than 30 times, and we could speculate that the cells at the 1$^{st}$ differentiation level should also have a higher proliferation potential (also much more than 30 divisions) than the theoretical potential. Therefore, the number of tumor cells theoretically capable of generating a new tumor would be more than the calculated number as described above.

### *3.2. A large population of tumor cells in B16 solid melanoma was potentially tumorigenic*

To further evaluate whether the tumor formation depends on the existence of CSCs, we needed to detect the potential tumorigenicity of any tumor cell in a tumor and examine if the number of tumor cells capable of generating a new tumor is higher than the number of CSCs. First we performed *in vitro* single-cell culture of tumor cells from murine B16 solid melanoma. We found that most of the B16 tumor cells died in a few days, while some formed small colonies but stopped proliferating after several days, indicating that they were possibly composed of terminally differentiated cells (Fig. 2, a and b). Meanwhile, there were large colonies containing ＞1000 cells formed in 2 weeks from 17.7% of the initial cultured B16 tumor cells (Fig. 2, c). The large colonies continued to proliferate after passaging. When the number of passaged cells from each large colony reached about 10$^6$, we inoculated immunocompetent syngeneic C57BL/6 mice with the tumor cells from some randomly selected colonies at a dose of $7.5 \times 10^5$ or $1.5 \times 10^5$ cells per mouse. We found that all the mice (13/13) developed melanoma at these inoculation doses, and all the tumors had diameters greater than 1 cm within 22 days (Fig. 3). When the tumors could be palpated, the tumor growth rate was similar between the 2 groups of mice inoculated with $7.5 \times 10^5$ or $1.5 \times 10^5$ cells, although the onset of tumor formation was delayed in the mice inoculated with $1.5 \times 10^5$ cells (Fig. 3, b). However, when the mice received only 750 cells from the same colonies, only 2 of 7 mice developed melanoma in 100 days (Fig. 3, a), indicating that the capacity of tumor generation from a tumor cell colony depends mainly on the number of inoculated cells and that a small number of cells may not be capable of adapting and proliferating *in vivo* in order to form a tumor. In fact, in the previous studies that identified CSCs (4, 5, 7, 15), inoculation with at least 10–100 or more tumor cells and not just 1 CSC was required to generate a new tumor, although theoretically a tumor could be generated after inoculation with a single CSC. Taken together, these results indicate that if only a few tumor cells, even if they are CSCs, are used for inoculation, they can only barely generate a tumor. Furthermore, more importantly, our results indicate that there are 17.7% of tumor cells in B16 melanoma with the potential to generate a new melanoma. Under suitable growth conditions (*in vitro* culture condition in these experiments) that support the proliferation of these single tumor cells to achieve a certain cell number, these cells can generate a tumor *in vivo*. Thus, similarly, if the microenvironment of any *in vivo* system is suitable for the growth of these single tumor cells, they may proliferate and generate a new tumor within that environment. Obviously, all the 17.7% of B16 tumor cells might have a high proliferation potential that they could divide >30 times and thus form a tumor *in vivo*. These 17.7% of B16 tumor cells might encompass the CSCs of B16 tumor (differentiation level 0, Fig. 1) and some of their offspring cells (differentiation level 1 and more, Fig. 1). According to the idealized model in Fig. 1 and the actual proportion of CSCs in a tumor (for example, $1/10^5$), the cells at the 1st differentiation level in B16 melanoma should have a proliferation potential that they can undergo 43–56 divisions (If there is only one CSC in a 1-cm$^3$ tumor, then $2^{y-29} = 0.177$, $y \approx 26$, and the proliferation potential is $30 + y = 56$; if the proportion of CSC is $1/10^5$, then there are $10^4$ ($10^4 \approx 2^{13}$) CSCs in a 1-cm$^3$ tumor, and the proliferation potential of the 1st differentiation level is $56 - 13 = 43$).

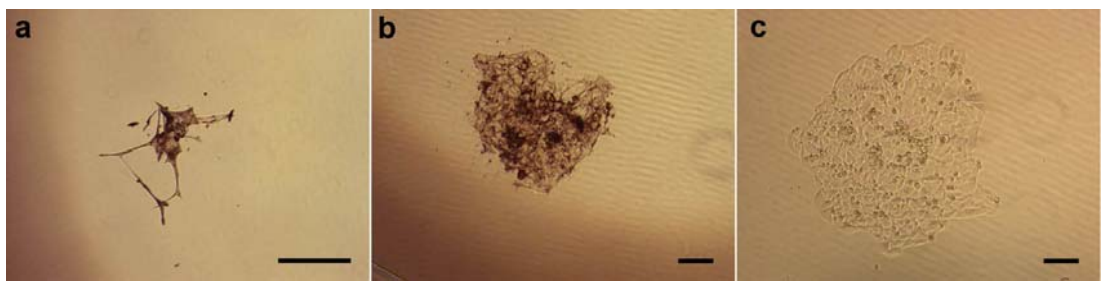

**Figure 2. Colony culture of B16 melanoma cells *in vitro*.** Single tumor cells from a murine B16 solid melanoma were cultured *in vitro* for 2 weeks, and 3 representative photographs of clonal colonies of different sizes are shown. Bar: 50 μm (a) and 100 μm (b and c).

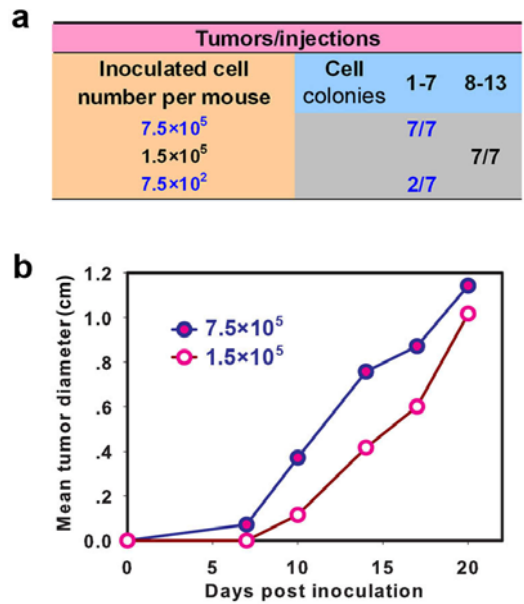

**Figure 3. Tumor formation of B16 melanoma cells *in vivo*.** Mice were inoculated with tumor cells from different clonal colonies at the indicated doses and tumor formation was monitored for 100 days (**a**). The individual tumor diameter was measured twice a week and the mean tumor diameters of two groups were shown in **b**.

### *3.3. A large population of tumor cells in either Lewis lung carcinoma or H22 hepatoma was potentially tumorigenic*

Next, we evaluated whether other types of tumors have characteristics similar to that of B16 melanoma. We performed similar experiments using *in vitro* single-cell cultures and tumor cell inoculation experiments for murine Lewis lung carcinoma (LLC) and H22 hepatoma. We found that 5.2% of the LLC solid tumor cells could form large clonal colonies *in vitro* (data not shown). Further, the tumor cells from these colonies could generate new tumors *in vivo* when inoculated at high doses into immunocompetent syngeneic mice (Fig. 4, a), similar to the results from B16 melanoma. As for H22 solid tumor cells, we found that 38.6% of them could form large clonal colonies *in vitro*. Also, the tumor cells from these colonies could generate new tumors *in vivo* when inoculated at relatively high doses (Fig. 4, b). Taken together, these results further suggested that the tumorigenic tumor cells is much higher than the number of CSCs, although the percentage of the tumorigenic tumor cells was varied among different tumor types.

### 4. Discussion

In this study, our results suggest that the number of tumor cells capable of generating a new tumor is higher than the number of CSCs. This was proved both in the theoretical calculation and in the 3 types of tumors we examined in our experiments. CSCs are a rare population, and they generally have higher tumorigenicity than other tumor cells, for example, they can self-renew and thus can preserve the stem cell properties even after serial *in vivo* passage. Our results indicate that CSCs are not essential for tumor formation. Besides CSCs, more other tumor cells could generate a new tumor, even they could probably not self-renew. The difference between CSCs and other tumorigenic tumor cells is that CSCs can self-renew, and thus after serial *in vivo* passage, there are still CSCs in the tumor which can 're-initiate' a tumor, while other tumorigenic tumor cells might only generate one generation of tumor and probably no tumorigenic tumor cells exist in the generated tumor.

a

| Inoculated cell number per mouse | Cell colonies | |
|---|---|---|
| | 1-6 | 7-12 |
| $7.5 \times 10^5$ | 6/6 | |
| $1.5 \times 10^5$ | | 6/6 |
| $7.5 \times 10^2$ | 0/6 | |

b

| Inoculated cell number per mouse | Cell colonies | | |
|---|---|---|---|
| | 1-2 | 3-8 | 9-14 |
| $3 \times 10^5$ | 2/2 | | |
| $1 \times 10^5$ | | 6/6 | |
| $2 \times 10^4$ | | | 2/6 |
| $1 \times 10^2$ | | 0/6 | |

**Figure 4. Tumor formation by LLC and H22 tumor cells in an *in vivo* system.** Mice were inoculated with LLC tumor cells (a) or H22 tumor cells (b) from different clonal colonies at the indicated doses. The mice were examined twice a week for tumors by observation and palpation for 18–100 days. The number of tumors that formed and the number of injections that were administered are indicated for each dose.

Our results are in agreement with the results of two recent studies. One of them showed that nonirradiated congenic mice transplanted with as few as 10 or even 1 unfractionated lymphoma cell developed tumors (8, 16), the other showed that more than one-fourth of the unselected melanoma cells from patients were tumorigenic (9). We speculate that in those studies even if CSCs were not present among the transplanted cells, but the transplanted cells had a high proliferation potential, i.e., they could undergo at least 30 divisions.

Our results have important implications for tumor therapy, although more studies need to be done to prove our conclusion, for example, in other kinds of tumors, especially in human tumors. CSCs might be more resistant to chemotherapy than more differentiated cancer cells, and if they survive, they will continue the process of tumor growth and progression, or generate new tumors, termed recurrence or metastasis (1). Our results suggest that new tumors may be generated from not only CSCs, but any tumor cells with a high proliferation potential. Since we do not know whether there are any locations in the body where the microenvironment is suitable for the growth of such tumorigenic cells, just targeting the CSCs might not be sufficient. We should consider all tumor cells that have a high proliferation potential. During therapy, either all these tumor cells including CSCs should be eliminated or microenvironments that support the proliferation of these tumor cells should be destroyed.

**Acknowledgments**

We acknowledge the assistance of AH Wang in data analysis and manuscript preparation. This study was jointly supported by China Postdoctoral Science Foundation, Scientific Research Fund of Shanghai Health Bureau (2008131) and Starting Fund of Fudan University for New Teachers.